\documentclass[
  aps,  
  twocolumn,
  prl,
  longbibliography,
  preprintnumbers,
  superscriptaddress,
  floatfix,
]{revtex4-2}
\usepackage{amssymb,amsfonts,amsmath,dsfont}
\usepackage{graphicx} 
\usepackage{gensymb}
\usepackage{pifont}
\usepackage{bm} 
\usepackage{color}
\usepackage{accents}
\usepackage{mathtools}
\usepackage{hyperref}
\usepackage{url}
\hypersetup{colorlinks=true, citecolor=blue, urlcolor=blue, linkcolor=blue}
\usepackage{multirow}

\makeatletter
\newcommand*{\supplementarystart}{%
  \close@column@grid%
  \clearpage%
  \onecolumngrid%
  \setcounter{enumiv}{0} % resets counter for references
  \setcounter{equation}{0} % resets counter for equations
  \setcounter{figure}{0} % resets counter for figs
  \setcounter{table}{0} % resets counter for tables
  \setcounter{page}{1}
  \c@secnumdepth=4
  \renewcommand{\theequation}{s\arabic{equation}} % equations numbered with S...
  \renewcommand{\bibnumfmt}[1]{[s##1]} % bibtems [S...]
  \renewcommand{\@onlinecite}{s\citealp} % citations [S...]
  \renewcommand{\cite}[1]{{[}\onlinecite{##1}{]}}
  \renewcommand{\thefigure}{s\arabic{figure}}
  \renewcommand{\thetable}{s\Roman{table}}
  \renewcommand{\thepage}{s\arabic{page}}
}
\makeatother
 
\newcommand{\s}{\sum\limits} 
 
\newcommand{\pa}{\partial} 
 
\newcommand{\be}{\begin{equation}} 
\newcommand{\e}{\end{equation}} 
\newcommand{\beml}{\begin{subequations}} 
\newcommand{\eml}{\end{subequations}} 
\newcommand{\beq}{\begin{eqnarray}} 
\newcommand{\eq}{\end{eqnarray}} 
\newcommand{\ba}{\begin{array}} 
\newcommand{\ea}{\end{array}} 
\newcommand{\bpm}{\begin{pmatrix}} 
\newcommand{\epm}{\end{pmatrix}} 
\newcommand{\bc}{\begin{cases}} 
\newcommand{\ec}{\end{cases}} 
\newcommand{\lt}{\left} 
\newcommand{\rt}{\right} 
\newcommand{\n}{\nonumber} 
\newcommand{\la}{\langle} 
\newcommand{\ra}{\rangle} 
\newcommand{\ep}{\varepsilon} 
 
\newcommand{\bs}{\boldsymbol} 
\newcommand{\bb}{\mathbf} 
\newcommand{\h}{^\dagger} 
\newcommand{\ph}{^{\phantom{\dagger}}}

\begin{document}

\title{Nonequilibrium magnons from hot electrons in antiferromagnetic systems}

\author{Marion M. S. Barbeau}
\affiliation{Center for Quantum Spintronics, Department of Physics, Norwegian University of Science and Technology, NO-7491 Trondheim, Norway}
\author{Mikhail Titov}
\affiliation{Radboud University, Institute for Molecules and Materials, 6525 AJ Nijmegen, The Netherlands}
\author{Mikhail I. Katsnelson}
\affiliation{Radboud University, Institute for Molecules and Materials, 6525 AJ Nijmegen, The Netherlands}
\author{Alireza Qaiumzadeh}
\affiliation{Center for Quantum Spintronics, Department of Physics, Norwegian University of Science and Technology, NO-7491 Trondheim, Norway}

\date{\today}

\begin{abstract}
We describe a \emph{nonthermal} magnon activation mechanism in antiferromagnetic (AFM) systems via locally equilibrated \emph{spin-unpolarized} hot electrons excited by an ultrafast intense laser pulse. We employ a quantum kinetic equation that takes into account a direct electron-magnon scattering channel in either bulk AFM metal or at the interface of the AFM/normal-metal heterostructure. The mechanism is responsible for the nonequilibrium population of AFM magnon modes on a subnanosecond timescale, which are formed shortly after quasithermalization of hot electrons by Coulomb interactions. Nonequilibrium magnon populations can be additionally manipulated by applying an external magnetic field. Our work paves the way toward spin dynamics control in AFM systems via the ultrafast manipulation of out-of-equilibrium magnon excitations.
\end{abstract}

\maketitle

%%%%%%%%%%%%%%%%%%%%%%%%%%%%%%%%%%%%%%%%%%%%%%%%
%%%%%%%%%%%%%%%%%%%%%%%%%%%%%%%%%%%%%%%%%%%%%%%%

The discovery of ultrafast spin control by subpicosecond laser pulses promises new opportunities for data storage and processing  \cite{Stanciu2007,Kirilyuk2010,Lambert2014}. At the same time ultrafast magnetization dynamics involve nonequilibrium phenomena that lack comprehensive {\em ab initio} or model description. The field has no shortage of experimental observations: ultrafast demagnetization and magnetic switching as well as nonequilibrium magnon population excitation.

The observed phenomena have obtained qualitative interpretations based on the mechanisms of laser-induced heating \cite{Beaurepaire1996,Ostler2012}, magnetic exchange engineering \cite{Mikhaylovskiy2015,Mentink2017,Losada2019}, light-induced fields \cite{Kimel2007,IFE1,IFE2}, etc.

Ultrafast photophysical phenomena are often successfully described within the phenomenological multitemperature model that assumes thermalized distributions of electron, magnon, phonon and other subsystems \cite{Kirilyuk2010}. Many experiments with conducting magnets assume initial excitation of nonthermal hot electrons and their energy transfer into thermalized spin and lattice degrees of freedom \cite{Matsubara_2015,SciRep2020}. Such energy transfer may, however, prompt nonequlibrium magnon excitations on transient timescales of electron-magnon interactions \cite{Tveten2015, Brener2017, Duine}. In particular, magnons in half-metal ferromagnets may be nonthermally excited by laser-induced hot electrons via nonquasiparticle (incoherent) states \cite{RevModPhys.80.315,Irkhin_2015, Brener2017, PhysRevLett.121.257201}. 
In half-metal ferromagnets, the electrons are almost fully spin polarized; hence, magnon excitations are facilitated by some virtual scattering processes \cite{Brener2017}. 

In this Letter, we consider antiferromagnetic (AFM) systems, where electron subbands are spin degenerate. 
We show how nonequilibrium magnons are nonthermally activated via a direct magnon-electron scattering channel. We assume that the strongly excited electron subsystem, by an ultrafast laser pulse, is initially quasithermalized at some hot temperature. 

Despite the vast literature on different aspects of electron-magnon interactions and the diversity of correlated materials with AFM ordering, there exists only a handful of studies of AFM magnon excitation by electrons \cite{Yu2020,Cheng2020}. 

Typical resonance frequency in AFM systems is in the terahertz range. This is in contrast to ferromagnetic (FM) systems where it can hardly exceed a few gigahertz. This property has been put forward as the basis for ultrafast AFM memory and computational devices \cite{Marti2014}. Unlike FM systems, the AFM materials are typically characterized by two circularly polarized magnon modes with opposite polarizations \cite{Rezende2019}. Energy transfer from hot electrons may induce a nonequilibrium population of these modes on transient time-scales. We argue that such nonequilibrium magnon kinetics is generic for AFM metals \cite{MetallicAFMs,PhysRevLett.120.197202} and AFM-insulator/normal-metal hetero-structures.

Below we consider an AFM semiconductor with two ($A$ and $B$) sublattices characterized by magnetic moments $\bb{S}_{Ai}$ and $\bb{S}_{Bi}$, where the index $i$ numerates magnetic unit cells. The AFM dynamics is described by the N\'{e}el vector $\bb{L}=\sum_{i=1}^{N/2}(\bb{S}_{Ai}-\bb{S}_{Bi})$, where $N$ is the total number of lattice sites. The value of localized magnetic moments $S$ is assumed to be sufficiently large to justify the expansion in a small parameter $(2S)^{-1}$.

Starting from an effective \emph {s--d(f)} model, we formulate a Boltzmann kinetic equation \cite{Frohlich1964,Holland1965,Nakata2021,Brener2017} for nonequilibrium magnons in the presence of thermalized hot electrons.

%%%%%%%%%%%%%%%%%%%%%%%%%%%%%%%%%%%%%%%%%%%%%%%%
%%%%%%%%%%%%%%%%%%%%%%%%%%%%%%%%%%%%%%%%%%%%%%%%

The model consists of an effective tight-binding Hamiltonian $H_\textrm{e}$ for conduction electrons that is coupled to an AFM Heisenberg model $H_\textrm{m}$ for localized spins by a local exchange interaction,
\be
\label{sdcop}
H_\textrm{sd} = - I_{\rm sd} \sum_{i=1}^N \bb{S}_i\cdot  c\h_{i\sigma} \bs{\sigma}_{\sigma\sigma'}  c\ph_{i\sigma},
\e
where $I_{\rm sd}$ is the \emph {s--d(f)} exchange parameter, $\bs{\sigma}$ is the vector of Pauli matrices, and $c_{i\sigma}(c\h_{i\sigma})$ is the corresponding electron annihilation (creation) operator at site $i$ with spin-$1/2$ index $\sigma$ \cite{vonsovsky1974magnetism,Yosida_Magnetism}.

We diagonalize the electronic Hamiltonian $H_\textrm{e}$ and linearize the magnon Hamiltonian $H_\textrm{m}$ with respect to its ground state [see the Supplemental Material (SM) \cite{SM}]. The resulting model reads $H=H_\textrm{e}+H_\textrm{m} + H_\textrm{sd}$, where
\beml
\label{model_em}
\begin{align}
H_\textrm{e}=&\s_{\bb{k}\sigma}\ep_{\bb{k}}\big[c\h_{c\bb{k}\sigma}c\ph_{c\bb{k}\sigma}- c\h_{v\bb{k}\sigma}c\ph_{v\bb{k}\sigma}\big],\\
H_\textrm{m}=&\s_{\bb{q}}\lt[\omega_{\bb{q}}^\alpha\alpha\h_{\bb{q}}\alpha\ph_{\bb{q}}+ \omega_{\bb{q}}^\beta\beta\h_{\bb{q}}\beta\ph_{\bb{q}}\rt],
\end{align}
\eml
represent the bosonic and fermionic sectors of the model, respectively \cite{PhysRevB.62.5647,PhysRevB.52.6181,Katsnelson91,PhysRevB.41.11457,Irkhin89,nonlinear-interactions}. 

The fermionic annihilation (creation) operators, $c_{m\bb{k}\sigma}(c\h_{m\bb{k}\sigma})$, are defined for conduction, $m=c$, and valence, $m=v$, bands with the dispersion $\ep_{m\bb{k}}=\pm \ep_{\bb{k}}$. Bosonic annihilation (creation) operators $\alpha_{\bb{q}} (\alpha\h_{\bb{q}})$ and $\beta_{\bb{q}}(\beta\h_{\bb{q}})$ refer to the two magnon branches with the dispersions $\omega_{\bb{q}}^{\alpha(\beta)}=\omega_{\bb{q}} \mp \Delta$ that are split by a Zeeman energy $\Delta$. The summations over $\bb{k}$ and $\bb{q}$ extend over the first Brillouin zone that is defined with respect to the magnetic lattice (with the double lattice spacing).

We consider a single-orbital tight-binding model of conduction electrons on a bipartite lattice with nearest-neighbor hopping $t_a$ and a Heisenberg model of collinear AFM on the same lattice \cite{PhysRevB.62.5647,vonsovsky1974magnetism,akhiezer1968spin,Yosida_Magnetism}, hence 
\be
\label{disp0}
\ep_{\bb{k}}=\sqrt{S^2 I^2_\textrm{sd}+z^2t_a^2|F_{\bb{k}}|^2}, \;
\omega_{\bb{q}}=2JzS\sqrt{1-|F_{\bb{q}}|^2},
\e
where $J$ is the Heisenberg exchange energy, $z$ is the coordination number, and $F_{\bb{k}}=z^{-1}\sum_{\alpha=1}^z\exp(i\bb{k}{\cdot}\bs{\delta}_\alpha)$ is the structure factor, where the vectors $\bs{\delta}_\alpha$ are the translation vectors to the nearest neighbor sites. In our model, the AFM electron band gap is set by $2S|I_{\rm sd}|$. 

We restrict ourselves to the lattice with the inversion symmetry and neglect spin-orbit interactions. For the sake of definiteness, we set the Fermi level $\ep_\textrm{F}$ at the conduction band and assume the limit $\ep_\textrm{F}\gg \Delta$. In this case the effect of external magnetic field can be taken into account in the form of Zeeman splitting of magnon modes only, while the splitting of conduction electron bands and, hence, spin polarization of itinerant electrons can be disregarded.  We also neglect nonlinear magnon-magnon interactions \cite{nonlinear-interactions}.

In the linear order with respect to $I_\textrm{sd}$ the electron-magnon interaction of Eq.~(\ref{sdcop}) can be rewritten as,
\begin{align}
&H_{\rm sd}=-\frac{\sqrt{S}I_{\rm sd}}{\sqrt{N}}\sum_{m,n\in\{c,v\}} \sum_{\bb{k},\bb{q}}\Big\{ \mathcal{V}_{mn}^{\alpha\uparrow}\alpha^{\dagger}_{\bb{q}}c^{\dagger}_{m\bb{k}\uparrow}c^{\phantom\dagger}_{n\bb{k}+\bb{q}\downarrow}\nonumber\\&+\mathcal{V}_{mn}^{\alpha\downarrow}\alpha^{\phantom\dagger}_{\bb{q}}c^{\dagger}_{m\bb{k}\downarrow}c^{\phantom\dagger}_{n\bb{k}-\bb{q}\uparrow}+\mathcal{V}_{mn}^{\beta\downarrow}\beta^{\dagger}_{\bb{q}}c^{\dagger}_{m\bb{k}\downarrow}c^{\phantom\dagger}_{n\bb{k}+\bb{q}\uparrow}\nonumber\\&+\mathcal{V}_{mn}^{\beta\uparrow}\beta^{\phantom\dagger}_{\bb{q}}c^{\dagger}_{m\bb{k}\uparrow}c^{\phantom\dagger}_{n\bb{k}-\bb{q}\downarrow}\Big\},
\label{s-d Hamiltonian}
\end{align}
where the dimensionless quantities $\mathcal{V}^{\gamma\sigma}_{mn}$ parametrize the inter- ($m \neq n$) and intra- ($m = n$) band transfer rates that are defined in the SM \cite{SM}.

%%%%%%%%%%%%%%%%%%%%%%%%%%%%%%%%%%%%%%%%%%%%%%%%

We apply the model of Eqs.~(\ref{sdcop}) and (\ref{model_em}) to describe magnon dynamics in a nonequilibrium situation: the initial state is formed by hot electrons, quasithermalized by Coloumb interactions at an effective temperature $T$, and by a negligible number of thermal magnons. Such a state is formed by a femtosecond laser pulse on picosecond timescales. The next nanosecond is dominated by energy transfer from an electron to a magnon subsystem due to the interaction of Eq.~(\ref{s-d Hamiltonian}). This is the process that we aim to describe with the kinetic approach. We find that magnon kinetics cannot be merely reduced to heating. It is characterized instead by an anomalous excitation of magnons with large momenta. 

In order to construct a kinetic equation for magnon densities, we employ a perturbation formalism that was originally developed by Fr\"{o}hlich and Taylor \cite{Frohlich1964} in the context of electron-phonon interaction. 

We use the Heisenberg picture to introduce time-dependent magnon densities,
\be
N_{\bb{q}}^{\alpha}(t)=\lt\la \alpha\h_{\bb{q}}(t)\alpha\ph_{\bb{q}}(t)\rt\ra,\quad 
N_{\bb{q}}^{\beta}(t)=\lt\la \beta\h_{\bb{q}}(t)\beta\ph_{\bb{q}}(t)\rt\ra,
\e
and interband magnon transition probabilities,
\be
P_{\bb{q}}^{\alpha\beta}(t)=\lt\la \alpha\h_{\bb{q}}(t)\beta\ph_{\bb{q}}(t)\rt\ra,\quad 
P_{\bb{q}}^{\beta\alpha}(t)=\lt\la \beta\h_{\bb{q}}(t)\alpha\ph_{\bb{q}}(t)\rt\ra,
\e
where the angular brackets represent the averaging over the canonical ensemble at an initial moment of time. Our goal is to derive a kinetic equation on these quantities to model their time evolution. 

Using nonequilibrium field theory \cite{SM}, we can cast these equations in the following form:
\beml
\label{kinetics}
\begin{align}
\pa_t N^{\gamma}_{\bb{q}}(t)
= &\lt[1-\big({e^{\frac{\omega^\gamma_{\bb{q}}}{k_{\text{B}}T}}-1}\big)N_{\bb{q}}^{\gamma}(t)\rt]\mathcal{I}_{\bb{q}}^{\gamma\gamma}(t), \label{Naa}\\
\partial_t P_{\bb{q}}^{\alpha\beta}(t)=&P_{\bb{q}}^{\alpha\beta}(t)\mathcal{I}^{\alpha\beta}_{\bb{q}}(t)-(1+P_{\bb{q}}^{\alpha\beta}(t))\mathcal{I}^{\beta\alpha}_{\bb{q}}(t),\label{QB equation Alpha Beta}\\
\partial_t P_{\bb{q}}^{\beta\alpha}(t)=&(1+P_{\bb{q}}^{\beta\alpha}(t))\mathcal{I}^{\alpha\beta}_{\bb{q}}(t)-P_{\bb{q}}^{\beta\alpha}(t)\mathcal{I}^{\beta\alpha}_{\bb{q}}(t),\label{QB equation Beta Alpha}
\end{align}
\eml
where $T$ is the effective temperature of hot electrons, $k_{\text{B}}$ is the Boltzmann constant, $\mathcal{I}^{\gamma{\gamma}}_{\bb{q}}$ and $\mathcal{I}^{\gamma\bar{\gamma}}_{\bb{q}}$ represent intra- and interband collision integrals, $\gamma=\{\alpha,\beta\}$, and $\bar{\gamma} \neq \gamma$.
Intraband collision integrals are defined as
\beml
\label{coll}
\begin{align}
\mathcal{I}^{\alpha\alpha}_{\bb{q}}= &\frac{2\pi SI_\textrm{sd}^2}{N} \sum_{m,n,\bb{k}}|\mathcal{V}^{\alpha\uparrow}_{mn}|^2\big[1-f(\ep_{m\bb{k}})\big]f(\ep_{m\bb{k}}+\omega^{\alpha}_{\bb{q}})\n\\
&\qquad\qquad \times\mathcal{A}^n_{\bb{k}+\bb{q}\downarrow}(t,\ep_{m\bb{k}}+\omega^{\alpha}_{\bb{q}}),\label{I Alpha}\\
\mathcal{I}_{\bb{q}}^{\beta\beta}= & \frac{2\pi S I_\textrm{sd}^2}{N} \sum_{m,n,\bb{k}} |\mathcal{V}^{\beta\downarrow}_{mn}|^2 \big[1-f(\ep_{m\bb{k}})\big]f(\ep_{m\bb{k}}+\omega^{\beta}_{\bb{q}})\n\\
&\qquad\qquad \times\mathcal{A}^n_{\bb{k}+\bb{q}\uparrow}(t,\ep_{m\bb{k}}+\omega^{\beta}_{\bb{q}}),\label{I Beta}
\end{align}
\eml
while the interband collision integrals $\mathcal{I}^{\gamma\bar{\gamma}}_{\bb{q}}(t)$ are given by,
\beml
\label{I_AB}
\begin{align}
\mathcal{I}^{\alpha\beta}_{\bb{q}} = & \frac{\pi SI_{\rm sd}^2}{N} \s_{m,n,\bb{k}}|\mathcal{V}^{\alpha\uparrow}_{mn}|^2\Big[f(\ep_{m\bb{k}}+\omega^{\beta}_{\bb{q}})\mathcal{A}^n_{\bb{k}+\bb{q}\downarrow}(t,\ep_{m\bb{k}}+\omega^{\beta}_{\bb{q}})\n\\
&-f(\ep_{n\bb{k}+\bb{q}}-\omega^{\beta}_{\bb{q}})\mathcal{A}^m_{\bb{k}\uparrow}(t,\ep_{n\bb{k}+\bb{q}}-\omega^{\beta}_{\bb{q}})\Big],
\label{I A}\\
\mathcal{I}^{\beta\alpha}_{\bb{q}} = &\frac{\pi SI_{\rm sd}^2}{N} \s_{m,n,\bb{k}} |\mathcal{V}^{\beta\downarrow}_{mn}|^2\Big[f(\ep_{n\bb{k}+\bb{q}}-\omega^{\alpha}_{\bb{q}})\mathcal{A}^m_{\bb{k}\downarrow}(t,\ep_{n\bb{k}+\bb{q}}-\omega^{\alpha}_{\bb{q}})\n\\
& -f(\ep_{m\bb{k}}+\omega^{\alpha}_{\bb{q}})\mathcal{A}^n_{\bb{k}+\bb{q}\uparrow}(t,\ep_{m\bb{k}}+\omega^{\alpha}_{\bb{q}})\Big],
\label{I B}
\end{align}
\eml
where 
\be
\label{spectral function}
\mathcal{A}^m_{\bb{k}\sigma}(t,\ep)=\frac{1}{\pi}\frac{\Gamma^{\sigma}_{m\bb{k}}(t,\ep)/2}{(\ep_{m\bb{k}}-\ep)^2+\big(\Gamma^{\sigma}_{m\bb{k}}(t,\ep)/2\big)^2},
\e
is the electron spectral function. Here, the inverse quasiparticle lifetime $\Gamma^{\sigma}_{m\bb{k}}(t,\ep)$ is defined by the electron-magnon scattering in the adiabatic approximation. Its dependence on the evolution time $t$ originates in the time dependence of magnon density.  $f(\ep)=\big(e^{(\ep-\ep_\textrm{F})/{k_{\text{B}}T}}+1\big)^{-1}$ is the equilibrium Fermi-Dirac distribution function of hot electrons in a quasithermalized state with temperature $T$.

The scattering rate $\Gamma^{\sigma}_{m\bb{k}}$ can be represented as a sum intra- ($m=n$) and inter- ($m\neq n$) band contributions 
\be
\Gamma^{\sigma}_{m\bb{k}}(t,\ep)=\frac{2\pi S I_{\rm sd}^2}{N}\sum_{n,\bb{q},\gamma}\Gamma_{mn}^{\gamma\sigma}(\bb{k},\bb{q};t,\ep), \label{scattering}
\e
where $\Gamma_{mn}^{\gamma\sigma}(\bb{k},\bb{q};t,\ep)$ is a dimensionless scattering rate for an electron with spin $\sigma$ and momentum $\bb{k}$, while $\bb{q}$ is the transferred momentum to or from a magnon, 
%\beml
%\label{rates}
\begin{align}
& \Gamma_{mn}^{\alpha\uparrow}=|\mathcal{V}^{\alpha\uparrow}_{mn}|^2\big[N_{\bb{q}}^{\alpha}+f(\ep_{n\bb{k}+\bb{q}})\big]\delta\big(\ep+\omega^{\alpha}_{\bb{q}}-\ep_{n\bb{k}+\bb{q}}\big),\n\\%\label{Gamma 1}
& \Gamma_{mn}^{\alpha\downarrow}=|\mathcal{V}^{\alpha\downarrow}_{mn}|^2
\big[1\!+\!N_{\bb{q}}^{\alpha}\!-\!f(\ep_{n\bb{k}\!-\!\bb{q}})\big]\delta\big(\ep\!-\!\omega^{\alpha}_{\bb{q}}\!-\!\ep_{n\bb{k}\!-\!\bb{q}}\big),\n\\%\label{Gamma 2}\\
& \Gamma_{mn}^{\beta\uparrow}=|\mathcal{V}^{\beta\uparrow}_{mn}|^2\big[1\!+\!N_{\bb{q}}^{\beta}\!-\!f(\ep_{n\bb{k}\!-\!\bb{q}})\big]\delta\big(\ep\!-\!\omega^{\beta}_{\bb{q}}\!-\!\ep_{n\bb{k}\!-\!\bb{q}}\big),\n\\%\label{Gamma 3}\\
& \Gamma_{mn}^{\beta\downarrow}=|\mathcal{V}^{\beta\downarrow}_{mn}|^2\big[N_{\bb{q}}^{\beta}+f(\ep_{n\bb{k}+\bb{q}})\big]\delta\big(\ep+\omega^{\beta}_{\bb{q}}-\ep_{n\bb{k}+\bb{q}}\big).\n%\label{Gamma 4}
\end{align}
%\eml
In the absence of a magnetic field, $\Delta=0$,  one finds $N_{\bb{q}}^{\alpha}=N_{\bb{q}}^{\beta}$, $\Gamma_{mn}^{\alpha\downarrow}=\Gamma_{mn}^{\beta\uparrow}$, and  $\Gamma_{mn}^{\alpha\uparrow}= \Gamma_{mn}^{\beta\downarrow}$, which reflects the degeneracy of magnon bands. 

%%%%%%%%%%%%%%%%%%%%%%%%%%%%%%%%%%%%%%%%%%%%%%%%
%%%%%%%%%%%%%%%%%%%%%%%%%%%%%%%%%%%%%%%%%%%%%%%%
%%%%%% fig:Gamma_k
%%%%%%%%%%%%%%%%%%%%%%%%%%%%%%%%%%%%%%%%%%%%%%%%
\begin{figure}
\includegraphics[width=0.50\textwidth]{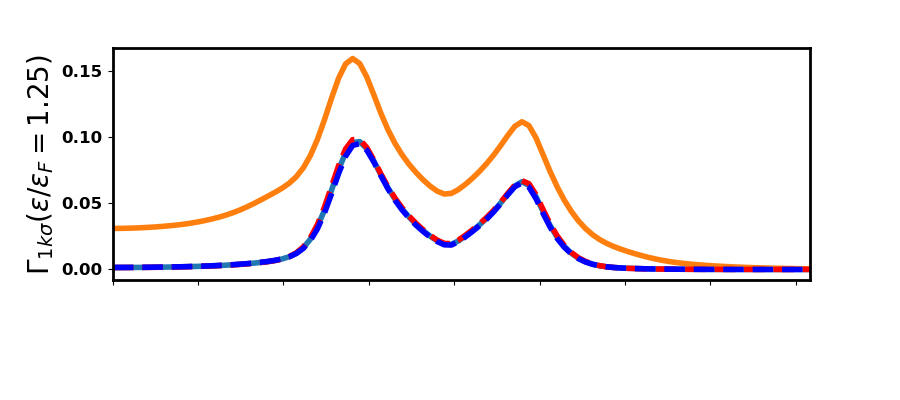}\\
\vspace{-1.2cm}
\includegraphics[width=0.50\textwidth]{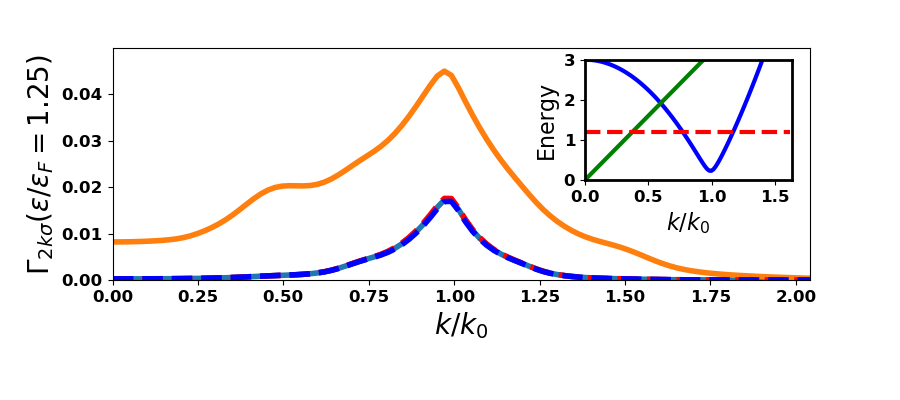}
\vspace{-1.2cm}
\caption{The scattering rate of hot electron quasiparticles in the conduction (top panel) and valence (bottom panel) bands at energy $\ep=1.5\,\ep_\textrm{F}$ from magnons as a function of the wavenumber. Solid lines correspond to $\Delta = 0$ for $k_{\text{B}}T = 0.2\,\ep_\textrm{F}$ (blue) and $k_{\text{B}}T = 0.4\,\ep_\textrm{F}$ (orange). Dashed lines show $\Gamma_{m\bb{k}}^{\uparrow}$ (blue) and $\Gamma_{m\bb{k}}^{\downarrow}$ (red) for $\Delta \sim \Delta_{\textrm{ani}}= 10^{-2}\ep_\textrm{F}$ and $k_{\text{B}}T= 0.2\,\ep_\textrm{F}$, which practically coincide with the solid blue line. We set $I_{\rm sd}=0.03 t_a$, $S=5/2$, and $\ep_\textrm{F}= 1.2\,t_a$. 
The inset shows the dispersion of electronic (blue) and magnonic (green) bands, in the unit of $t_a$, in the absence of a magnetic field, while the dashed red line in the inset corresponds to the Fermi energy.}
\label{fig:Gamma_k}
\end{figure}
%%%%%%%%%%%%%%%%%%%%%%%%%%%%%%%%%%%%%%%%%%%%%%%%%

The variation of the nonvanishing component of the N\'{e}el vector $\delta L(t)  = L_{\rm{z}}(t)-L_{\rm{z}}(0)$ is directly related to the magnon density $N^{\gamma}_{\bb{q}}(t)$ as
\be
\delta L(t) = \sum_{\textbf{q}}(|u_{\textbf{q}}|^2+|v_{\textbf{q}}|^2)\big(N^{\alpha}_{\bb{q}}(t)+N^{\beta}_{\bb{q}}(t)\big),
\e
where $u_{\textbf{q}}$ and $v_{\textbf{q}}$ are Bogoliubov coefficients, defined in the SM \cite{SM}.

%%%%%%%%%%%%%%%%%%%%%%%%%%%%%%%%%%%%%%%%%%%%%%%%
%%%%%%%%%%%%%%%%%%%%%%%%%%%%%%%%%%%%%%%%%%%%%%%%
%%%%%% fig:Nq
%%%%%%%%%%%%%%%%%%%%%%%%%%%%%%%%%%%%%%%%%%%%%%%%
\begin{figure}[t]
\includegraphics[width=0.5\textwidth]{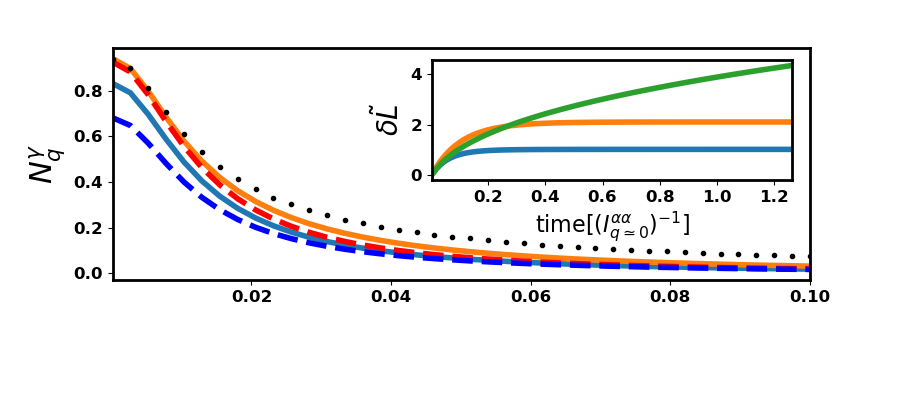}\\
\vspace{-1.2cm}
\includegraphics[width=0.5\textwidth]{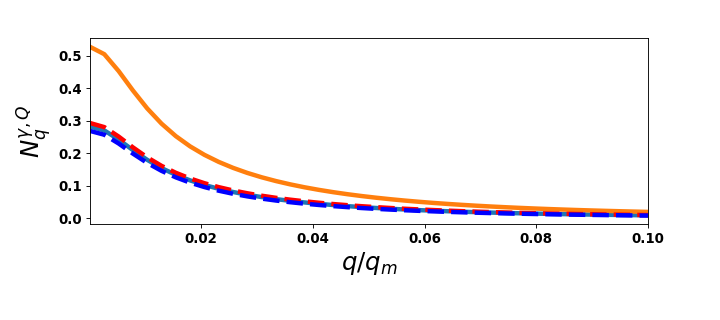}
\vspace{-1cm}
\caption{Nonequilibrium magnon density $N^{\gamma}_{\bb{q}}$ (top panel) and its quasiparticle approximation $N^{\gamma,Q}_{\bb{q}}$ (bottom panel) as a function of the magnon wavenumber at dimensionless time $t=0.85/I_{q\simeq0}^{\alpha\alpha}(k_{\text{B}}T=0.2\ep_\textrm{F},\Delta=0)$. Solid blue and solid orange lines show magnon distributions in the absence of a magnetic field where both magnon modes are degenerate at $k_{\text{B}}T=0.2\ep_\textrm{F}$ and $k_{\text{B}}T=0.4\ep_\textrm{F}$, respectively. Close to the spin-flop field, $\Delta=\Delta_{\textrm{ani}}{\simeq} 10^{-2}\ep_\textrm{F}$ and $k_{\text{B}}T=0.2\ep_\textrm{F}$, magnon distributions of the $\alpha$ and $\beta$ modes are presented by red and blue dashed lines, respectively. The parameters are the same as for Fig. \ref{fig:Gamma_k}. 
Black dotted curve in the top panel represents the thermal magnon number from the Bose-Einstein distribution for $k_{\text{B}}T=0.4\ep_\textrm{F}$ and $\Delta=0$. We divided the actual numbers into  $n^\gamma_{\bm{q}=0}\approx 40.8$ to make the curve in the same scale of the nonthermal magnons.
The inset shows the time evolution of the normalized N{\'e}el vector field, $\delta \tilde{L}=\delta L /(|u_{q=0}|^2+|v_{q=0}|^2)$. The blue and orange lines correspond to $k_{\text{B}}T=0.2\ep_\textrm{F}$ and $k_{\text{B}}T=0.4\ep_\textrm{F}$ in the absence of a magnetic field, respectively, while the green line corresponds to $k_{\text{B}}T=0.2\ep_\textrm{F}$ in the presence of a magnetic field close to the  spin-flop field. 
}\label{fig:Nq}
\end{figure}
%%%%%%%%%%%%%%%%%%%%%%%%%%%%%%%%%%%%%%%%%%%%%%%%%

Scattering rates $\Gamma_{m\bb{k}}^\sigma(t,\ep)$ do increase with time due to the fact that magnon densities are increasing. Such an increase has, in turn, a strong effect on electron-magnon collisions that become more probable. This phenomenon goes beyond quasiparticle approximation that assumes time-independent scattering rates. 

Due to the Lorentzian shape of the spectral function in Eq.~(\ref{spectral function}) the scattering rates $\Gamma_{i\bb{k}}^\sigma(t,\ep)$ make sense only in a vicinity of  the mass shell $\ep=\ep_{m\bb{k}}$. In our numerical analysis below we do, however, explicitly perform integration over energy. 

For the sake of illustration, we apply the kinetic theory developed above to an AFM metal on the cubic lattice assuming that the conduction electron Fermi energy is sufficiently far from the half filling (see the inset in the lower panel of Fig.~\ref{fig:Gamma_k}). For a stronger s-d(f) interaction close to the half-filling limit, see the SM \cite{SM}. 

To be more realistic, we assume a uniaxial easy-axis magnetic anisotropy of $K_z\sim 10^{-6}\,\ep_\textrm{F}$, that enters the AFM resonance frequency and opens up a band gap, $\Delta_{\textrm{ani}}\sim10^{-2}\,\ep_\textrm{F}$, in the magnon spectrum \cite{SM}. Magnon dispersion is also illustrated at the inset of Fig.~\ref{fig:Gamma_k} but the gap is too small to be visible. 

We perform a numerical simulation of magnon kinetics for two different temperatures of hot electrons: $k_{\text{B}}T=0.2\varepsilon_F$ and $k_{\text{B}}T=0.4\varepsilon_F$. We also illustrate the effect of Zeeman coupling $\Delta$ by comparing the case with two degenerate AFM magnons $\Delta=0$, and the case close to the AFM spin-flop transition $\Delta \sim \Delta_{\textrm{ani}}$ \cite{SM}.
 
In  Fig.~\ref{fig:Gamma_k} we plot electron-magnon scattering rates $\Gamma_{m\bb{k}}^\uparrow$ and $\Gamma_{m\bb{k}}^\downarrow$ shown with dashed blue and dashed red lines, respectively, as a function of the wavevector $\bb{k}$. The field $\Delta=\Delta_{\textrm{ani}}$ is still too small to induce any reasonable effect on the scattering rates. The contribution of the interband scattering to the total scattering rate is further reduced by increasing the Fermi energy and the electronic band gap.

The conduction electron scattering rate $\Gamma_{c\bb{k}}^\sigma$ has two peaks at two Fermi wavevectors, while the valence scattering rate $\Gamma_{v\bb{k}}^\sigma$ has a single peak at the bottom (top) of the conduction (valence) band.
 
We assume that at an initial moment of time $t=0$, quasiequilibrium hot spin-unpolarized electrons are described by the Fermi-Dirac distribution with an effective temperature $T$, while there is no nonequilibrium magnon $N^{\gamma}_{\bb{q}}(t=0)=0$ in the system. We then apply the Boltzmann equations of Eqs.~(\ref{kinetics}) to compute nonequilibrium magnon excitations for each magnon band, Eq.~(\ref{Naa}), and for the interband magnon scattering probabilities, Eqs.~ (\ref{QB equation Alpha Beta}) and (\ref{QB equation Beta Alpha}).

%%%%%%%%%%%%%%%%%%%%%%%%%%%%%%%%%%%%%%%%%%%%%%%%%%
%%%%%% fig:Pq
%%%%%%%%%%%%%%%%%%%%%%%%%%%%%%%%%%%%%%%%%%%%%%%%
\begin{figure}[t]
\includegraphics[width=0.5\textwidth]{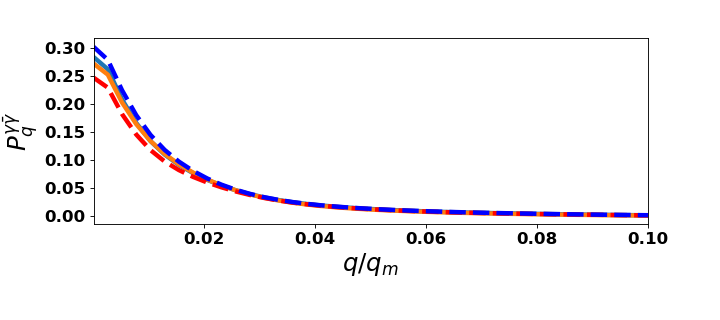}
\vspace{-1.3cm}
\caption{The interband transition probability between two magnon modes as a function of magnon wavenumber. Solid blue and solid orange lines show the scattering rate in the absence of a magnetic field, where interband scattering probabilities are degenerate at $k_{\text{B}}T=0.2\ep_\textrm{F}$ and $k_{\text{B}}T=0.4\,\ep_\textrm{F}$, respectively.
For $\Delta=\Delta_{\textrm{ani}}{\simeq}10^{-2}\,\ep_\textrm{F}$ and $T=k_{\text{B}}0.2\,\ep_\textrm{F}$, the interband scattering probabilities $P^{\alpha\beta}_{\bb{q}}$ and $P^{\beta\alpha}_{\bb{q}}$ are presented by red and blue dashed lines, respectively. The parameters are the same as those for Fig. \ref{fig:Nq}.}
\label{fig:Pq}
\end{figure}
%%%%%%%%%%%%%%%%%%%%%%%%%%%%%%%%%%%%%%%%%%%%%%%%%%%%%%%%%

In the absence of a magnetic field, both magnon modes are degenerate, and the number of magnons increases monotonously with temperature. In the presence of a magnetic field, the degeneracy of magnon modes is lifted. As a result, the number of left-handed $\alpha$-magnons becomes larger than the number of right-handed $\beta$-magnons, as shown in Fig. \ref{fig:Nq}. Since circularly polarized magnon modes carry spin angular momentum, the electron-magnon scattering generates a net nonequilibrium spin polarization in the system.
To compare magnon number distribution of nonthermally activated magnons and thermal magnons at hot electron temperature, we plot Bose-Einstein distribution $n^\gamma_{\bm{q}}=\big(e^{\omega^{\gamma}_{\bm{q}}/{k_{\text{B}}T}}-1\big)^{-1}$ in Fig. \ref{fig:Nq}. The number of thermal magnons at hot electron temperature is an order of magnitude larger than the nonthermally activated magnons. The effective temperature of nonequilibrium magnons is in the order of $I_{\rm sd}/2$. The wavenumber dependency of equilibrium and nonequilibrium magnons is different at large wavenumbers.
We do not expect a population of a significant amount of thermal magnons in the subpicosecond timescale.  

It is instructive to consider a ``classic'' limit for nonequilibrium magnon distribution $N^{\gamma}_{\bb{q}}$ by formally replacing the Lorentzian spectral function of Eq.~(\ref{spectral function}) with the Dirac delta function, $\mathcal{A}^{m,Q}_{\bb{k}\sigma}(\ep)= \delta(\ep_{m\bb{k}}-\ep)$. Such a quasiparticle approximation corresponds to the time-dependent Fermi golden rule. The corresponding magnon densities, $N^{\gamma,Q}_{\bb{q}}$, are plotted in the bottom panel of Fig. \ref{fig:Nq}. One can see that the quasiparticle approximation generally underestimates magnon densities.

Interband magnon transition probabilities between two AFM magnon modes are shown in Fig.~\ref{fig:Pq}. In the absence of a magnetic field, both interband probabilities are equal and decrease with temperature. In the presence of the field, the scattering probabilities $P^{\alpha \beta}_{\bb{q}}$ and $P^{\beta  \alpha}_{\bb{q}}$ become different and may, in principle, acquire a nonmonotonous temperature dependence.

The nonequilibrium AFM magnons can be activated via interfacial electron-magnon scattering \cite{Bender2012,Tveten2015,Manchon2017}. To describe such processes within our formalism, one should only replace the bulk $I_{\rm{sd}}$ in Eq.~(\ref{s-d Hamiltonian}) with the corresponding interface interaction.

Nonequilibrium magnons excited in the AFM layer can be detected by different methods, e.g., electrically by means of the inverse spin Hall effect in an AFM-heavy metal heterostructure or direct optical measurement of the AFM N{\'e}el vector.
The activation of the nonthermal magnons is equivalent to tilting the N{\'e}el vector from its equilibrium direction, which is a measurable quantity by optical techniques. The dynamics of the N{\'e}el vector $\delta L$ for different temperatures and in the presence of a magnetic field are plotted in the inset of Fig. \ref{fig:Nq}. In the absence of magnetic field, $\delta L$ increases but saturates very quickly after quenching (see blue and orange lines in the inset of Fig.~\ref{fig:Nq}), while in the presence of magnetic field, $\delta L$ increases continuously as a function of time (see green line in the inset of Fig.~\ref{fig:Nq}). It will be thermalized and saturated over longer timescales by magnon-phonon interactions.

In summary, we developed quantum Boltzmann equations for nonthermal AFM magnon densities in the presence of hot electrons. We showed that one may activate and control nonequilibrium AFM magnon populations with an external magnetic field. We discussed the qualitative importance of electron scattering on magnon evolution and compared our results to magnon quasi-classic approximation. Considering electron-magnon scattering rates beyond quasi-classical approximation, one suppresses the effects of temperature but enhances those of magnetic field. 
These results show that nonthermal magnon distribution can be activated by hot electrons in AFM systems.
Our results pave the way for the emerging field of AFM spintronics, AFM magnon condensation \cite{Volovik}, and ultrafast spin dynamics in correlated systems.

%%%%%%%%%%%%%%%%%%%%%%%%%%%%%%%%%%%%%%%%%%%%%%%%
%%%%%%%%%%%%%%%%%%%%%%%%%%%%%%%%%%%%%%%%%%%%%%%%
\section*{Acknowledgment}
The authors would like to thank S. Brener, B. Murzaliev, and R.E. Troncoso for insightful conversations. This project has been supported by the Norwegian Financial Mechanism Project No. 2019/34/H/ST3/00515, ``2Dtronics''; and partially by the Research Council of Norway through its Centers of Excellence funding scheme, Project No. 262633, ``QuSpin''.

\bibliography{apssamp.bib}

\end{document}